\documentclass[fleqn,twoside]{article}
\usepackage{espcrc2}

\usepackage{graphicx}
\usepackage[figuresright]{rotating}

\newcommand{\AmS}{{\protect\the\textfont2
  A\kern-.1667em\lower.5ex\hbox{M}\kern-.125emS}}

\title{Actions for dynamical fermion simulations: are we ready to go?}

\author{K. Jansen\address[MCSD]{NIC/DESY Zeuthen, 
        Platanenallee 6, 15738 Zeuthen, Germany}%
        \thanks{email: Karl.Jansen@desy.de}}
       
\begin{document}

\begin{abstract}
A critical review, playing devil's advocate, on present dynamical fermion
simulations is given. 
\vspace{1pc}
\end{abstract}

\maketitle

\vspace*{-0.4cm}
\section{Introduction}

No doubt, the lattice approach to quantum field theory has established itself
as an integral part of high energy theoretical physics and has produced 
many 
results that are relevant to understand
or even construct quantum field theories, as well as for phenomenology
helping to interpret experimentally obtained data.
Still, we are by far not at the end of the story. Serious dynamical fermion
simulations with realistic number of quarks, i.e. two light and one heavy
flavour, have just started and reveal already a number of unexpected
problems. It is the objective of this presentation to address these problems
and discuss their relevance and possible consequences. Nevertheless, 
we have made a lot of progress and we have understood much better
to keep under control 
systematic errors such as discretization and finite size effects, 
the chiral limit and non-perturbative renormalization.                
However, still more patience has to be invested in our
lattice computations to tell our colleagues from experiment trustworthy
numbers. New computer architectures are essential 
%, in particular of 
%the very ambitious QCDOC and APE projects as well as progress 
%with PC cluster systems 
and will help tremendously
to step forward.
Of course, we are all hoping for a miraculous invention of
a new algorithm that would solve all our problems at once,
although reality tells us that we will proceed in only small steps.

\vspace*{-0.1cm}
\section{Conceptual questions}

Certainly, for a non-lattice person (and even for a lattice practitioner)
the plethora of actions that are used in todays lattice 
simulations is, to say it moderately, confusing. The initial 
Wilson fermion action \cite{wilson} got many friends such 
as Symanzik improved actions 
\cite{symanzikfermion,symanzikquenched,symanzikdynamic} employing various 
gauge actions such as the Wilson \cite{wilson}, various forms of 
Symanzik improved gauge actions \cite{symanzikgauge} and renormalization 
group inspired actions \cite{rgaction}. 
A recent development concerning Wilson fermions is the formulation of 
twisted mass QCD \cite{tmqcd}. 
Of course, there is the approach 
of staggered fermions \cite{staggered} which comes again in improved forms 
\cite{staggeredimproved} and with different gauge actions. 
Newer developments are chiral invariant lattice fermions 
such as domain wall \cite{dw}, overlap 
\cite{overlap} and perfect actions \cite{truncated}. 
If exact chiral symmetry 
is not insisted on, 
but only chirally improved fermions are considered, 
{\em designer actions} such as
FLIC \cite{flic}, Hypercube \cite{hypercube} or truncated fixed point 
fermions 
\cite{truncated} enter the stage. 

For me, the criterion for choosing an action should be guided
by the demand that 
\begin{itemize}
\item {\em the results obtained with these actions 
agree among themselves in the continuum limit} and thus provide
valuable cross checks
\item {\em and if using a particular action will not waste computer time.}
\end{itemize}

There are a number of advantages and disadvantages of the above mentioned 
actions, 
concerning field theoretical conceptual aspects,
lattice artefacts, chiral and renormalization 
properties and simulation cost that 
will be addressed below. Let me start by discussing locality 
properties of the actions 
used.

\vspace*{-0.1cm}
\subsection{Rigorous actions}

The list of the above actions can be ordered by the level of rigor 
they obey in constructing a lattice quantum field theory with the 
emphasis on their locality properties.
The initial, standard Wilson actions 
(including the twisted mass formulation without
clover term) have been shown to be reflection positive 
\cite{positive}, they have a positive definite transfer matrix
\cite{transfermatrix} and thus obey the reconstruction theorem 
(see first two refs. in \cite{positive}) guaranteeing thus
the reconstruction of the
Minkowski Green's functions. 
Naive staggered fermions are on the same level \cite{staggeredrigor} with 
the peculiarity that the transfer matrix has to be constructed over
two lattice spacings.

\vspace*{-0.1cm}
\subsection{Non-rigorous but local actions}

The next class of actions are the ones that are ultra-local, i.e. 
they have an interaction range of only a few lattice points, but
for which reflection positivity cannot be proven. To this class
belong the Symanzik improved and the designer actions mentioned above.
However, there cannot be a serious doubt
that these actions are 
theoretically sound as the improvement terms added are irrelevant 
and vanish in the continuum limit. 
We might speak of ``physical positivity''.

\vspace*{-0.1cm}
\subsection{Exponentially localized actions}

The chirally invariant lattice actions even give up ultra-locality 
\cite{horvath,bietenholzultra} and only show an exponential 
localization \cite{locality}, as shown for domain wall 
\cite{dwlocal}, overlap \cite{locality} and perfect fermions
\cite{perfectlocal}. Again, we have no reason to doubt that these actions
define a local quantum field theory and 
describe QCD in the continuum limit. 

\vspace*{-0.1cm}
\subsection{Non-local actions (?)}

Every time a lattice Dirac operator is constructed, where the 
action ranges over all points on the lattice, the question 
of locality has to be asked anew. An example for such an operator
are improved 
staggered fermions for exact, say, two flavour simulations 
\cite{staggeredimproved}.
It is not clear, how such a staggered fermion operator can be
defined and also in this contribution no solution will be provided.
The suggestion that is used in {\em exact dynamical simulations}, 
is to take the square root of the 
staggered operator \cite{staggeredsquareroot}, implemented by some
polynomial approximation, which clearly 
defines an action that spreads over all lattice points.
It should hence be tested,
whether this operator is at least exponentially localized
(see my plenary talk \cite{myplenary} for such a test on a Wilson operator). 

There is a clear danger that through the square root operation a 
non-locality is introduced\footnote{There might be additional 
shortcomings of such a definition of a two flavour theory such as 
the violation of the optical theorem. However, it is not clear 
(at least not to me), how such a violation can be detected in practical
simulations.}.
%In the talk at the conference, F. Knechtli and myself have tested the 
%behaviour of the square root operator at the example of the hermitian
%operator $D^\dagger D = Q^2$ with $Q=\gamma_5 D$ and $D$ the standard 
%Wilson-Dirac operator. Taking the square root of $Q^2$ is non-trivial.
%As a solution of the equation $\sqrt{Q^2} = X$ we would only allow 
%a positive definite matrix $X$. Such a solution can be shown to be 
%unique \cite{uniquesqrt}. Therefore, if the operator $Q$ would only 
%have positive eigenvalues, the solution would be given by $X=Q$.
%This is not the case, however, as we explicitly checked at the 
%parameters we performed the simulations to check locality. 
%Hence the matrix $X$ is highly non-trivial. 

%We constructed $X$ by a polynomial approximation to $\sqrt{Q^2}$.
%Applying the so constructed operator to a test vector, a delta source,
%we checked the locality of the resulting operator as a function 
%of the taxi-driver system following very closely ref.~\cite{locality}.
For a fixed set of parameters, i.e. fixed pion mass and fixed coupling,
it should be expected that 
the localization of the operator is exponential, with couplings decaying
like $e^{-r/r_\mathrm{local}}$ with $r_\mathrm{local}$ the localization
range. In the continuum limit, 
staying at a fixed value of $r_0\cdot m_\pi$, in principle two 
outcomes are possible. In scenario I, 
$
r_\mathrm{local}\cdot m_\pi = \mathrm{constant}\;
\mathrm{for}\;\; a\rightarrow 0\; .
$
In this case we would obtain a continuum field theory with a localization
range of the order of the Compton wavelength of the pion. This would be
clearly un-acceptable. In order to save such 
a theory at least to some extent, it would be necessary that the 
localization 
range is much smaller than the Compton wavelength of the 
heaviest physical particle of the
target continuum field theory. 

In scenario II, 
$
r_\mathrm{local}\cdot m_\pi \rightarrow 0\;
\mathrm{for}\;\; a\rightarrow 0\; .
$
In this case we would get a local continuum quantum field theory. 
One may suggest a ``wishlist'', i.e. to 

\vspace*{0.3cm}
\fbox{
\parbox[b]{6.5cm}{
perform 
a check of the localization properties of the square root 
staggered fermion operator.
}
}
\vspace*{0.3cm}

Such a work is in progress \cite{staggeredcheck}. 
We assume first, by lack of an alternative, 
that the two flavour staggered operator 
$M_\mathrm{staggered}$ is defined by taking the square root of the
four flavor staggered operator.
%one considers $(M_\mathrm{staggered})^{1/4}$. 
Second, we think of an {\em exact} simulation which will 
employ this square root
operator. 
In particular, following the arguments for deriving a strict locality
bound of ref.~\cite{locality},  
we expect the localization range to 
satisfy the bound 
$r_\mathrm{local}\cdot m_\mathrm{quark}=\mathrm{constant}$. 
This behaviour follows directly from considering admissible
gauge field configurations, the knowledge of the maximum
eigenvalue of the staggered operator and the fact that the smallest 
eigenvalue is determined by the square of the bare quark mass. 
It remains to be seen, whether this theoretical bound is satisfied
in practical simulations or whether it grossly over-estimates
the localization range as it was found in the overlap case
\cite{locality}.

\vspace*{-0.1cm}
\subsection{Perturbation theory}

For most of the actions just discussed, despite their sometimes
complicated nature, perturbation theory has been worked out.
There are famous seminal papers for 
staggered fermions, see the 
first ref. in \cite{staggeredrigor} and \cite{shara}, 
as well as for Wilson fermions \cite{bochi}. 
These works build an important cornerstone to
understand many aspects of lattice field theory such as renormalization.
For more information on 
perturbation theory I refer to the recent review \cite{stefanopert}
and the talk by H. Trottier at this conference \cite{trottier}.

The only aspect of perturbation theory I would like to mention here
is the Reisz power counting theorem \cite{powercountreisz}. This theorem
guarantees that lattice integrals that appear in lattice perturbation
theory exist in the continuum limit. The theorem is satisfied for
Wilson fermions. However, staggered fermions do actually not obey
one of the conditions of the theorem (see \cite{powercountmartin}).
This does certainly not imply that there is a problem for staggered
fermions in perturbation theory. However, for a consistent and well 
founded perturbative discussion of staggered fermion it would be
very helpful to 

\vspace*{0.3cm}
\fbox{
\parbox[b]{6.5cm}{
construct a ``Reisz theorem'' for staggered fermions. 
}
}
\vspace*{0.3cm}

Such a theorem for staggered fermions might also be very useful to 
analyze and underpin improved staggered actions, possibly leading to 
a disentanglement of the tastes in perturbation theory.

%\subsection{C,P \& T}
%
%I would finally like to mention some works that discuss the study of 
%violations of CP and even CPT. M. Creutz \cite{creutz} discusses 
%how CP can be violated when the quark is chosen to be zero or 
%even negative. He finds interesting scenarios and the message 
%here in general is that aspects of physics at very small values of the
%quark mass might be rather different from what we know from the
%heavier quark mass values. Of course, the low energy phenomena will
%still be described by chiral perturbation theory at the same time. 
%
%In ref.~\cite{klinkhamer} a possible violation of CPT invariance 
%for overlap fermions is discussed. It is shown that CPT is violated
%for certain special gauge field configurations. It is not clear, however,
%whether these gauge fields form a set of measure zero and
%are hence not important. Further discussion on this point can be 
%found in \cite{susuki}.
%

\vspace*{-0.1cm}
\section{Algorithmic questions}

Algorithms are the backbone of simulations in lattice field theory
and testing and improving the algorithms cannot be appreciated 
enough. A review of existing algorithms for dynamical fermion simulations
can be found in \cite{peardon}. Promising new ideas 
can be found in \cite{annalgo}, \cite{hasenbuschidea} and 
\cite{martinalgo}, see
also \cite{hasenbusch}. 
%Developments using the 
%Schwartz alternating procedure to drive the global fermion algorithms
%more local are proposed in \cite{martinalgo}.   

The problem with our dynamical fermion simulations is the ``cost wall''.
We know to some extent the scaling behaviour of the algorithms and one
formula for the cost $C$ in the case of Wilson fermions 
according to ref.~\cite{ukawa} is 
\begin{equation}
C=A\cdot 
\left(\frac{m_\pi}{m_\rho}\right)^{-z_\pi}
L^{z_L}
a^{-z_a}\; .
\label{algocost}
\end{equation}
Here the exponents are given by 
$z_\pi=6$, $z_L=5$ and $z_a=7$; the value of the 
constant $A$ can be found in
\cite{ukawa}. 
Similar formulae are found by other groups \cite{panelalgo}.
Although this cost formula might still be plagued by a rather large 
uncertainty, it provides a good measure for the costs of the
algorithms employed.

\vspace*{-0.2cm}
\begin{figure}[htb]
\includegraphics[width=7.0cm,height=6.8cm]{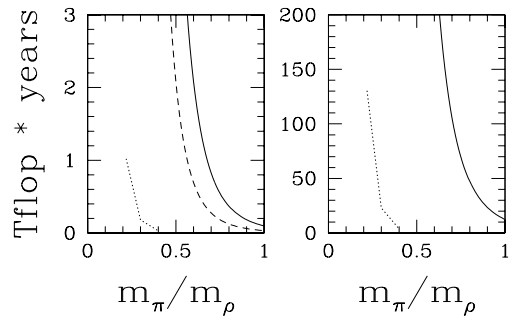}
\vspace*{-2.7cm}
\caption{A comparison of the cost of dynamical fermion simulations 
with (inexact) staggered (dotted line) and improved Wilson fermions 
(full line). 
The dashed line is the cost for Wilson fermions if the algorithms
would perform  a factor of four better than found in \cite{ukawa}.
The left plot is for a value of the lattice spacing $a=0.09$fm 
and the right plot for $a=0.045$fm.}
\label{fig:algocomp}
\end{figure}

In fig.~\ref{fig:algocomp} we show a comparison of simulations with 
staggered 
(inexact, i.e. using the $R$-algorithm \cite{gottliebhmc}) 
and Wilson fermions. Several things are remarkable in this 
graph. The first is that for both kind of fermions at some value of the quark
mass the costs explode. This is the famous cost wall where the simulations 
become essentially un-feasible. The second point is that for staggered
fermions this point is reached at much lower values of the quark mass
than for Wilson fermions. For the left plot we have used {\em measured}
cost data from improved staggered fermion simulations \cite{gottliebcost}
and the cost formula eq.~(\ref{algocost}) for Wilson fermions.
A third observation is that the functional dependence of the cost 
on the quark mass seems to be the same for both actions,
at least, given the accuracy of the
data used here. 
The right plot shows again a cost comparison at a lower value of the 
lattice spacing, exhibiting a frightening need of teraflop years. 
The apparent advantage of the simulation cost using improved staggered
fermions makes it even more important and urgent to test the theoretical
basis of this approach as discussed in section 2.

\begin{figure}[htb]
%\vspace{9pt}
\includegraphics[width=7.0cm,height=5.3cm]{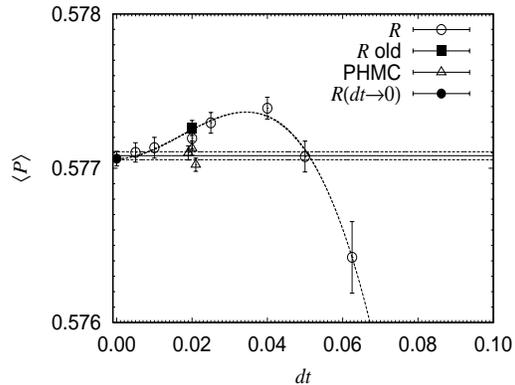}
\vspace*{-1.0cm}
\caption{A comparison of the plaquette from 
an exact odd flavour simulation (triangles) and 
an in-exact one (circles, square) against the step-size $dt$, 
from the first reference of \cite{oddflavour}.}
\label{exactvsinexact}
\end{figure}

One shortcoming of the staggered fermion simulations so far is that only 
a non-exact algorithm is used \cite{gottliebhmc}. In principle
we know how to execute an exact odd flavour algorithm 
\cite{oddflavour}. A comparison of the in-exact algorithm and the 
exact one is shown in fig.~\ref{exactvsinexact}. 
A striking effect of this graph is that the extrapolation to zero
step size as needed in the in-exact simulation is non-monotonic.
Hence such an extrapolation has to be done with very great care
using many simulation points. Since exact odd flavour algorithms are
on the market, it should be urged that they are indeed used in the 
staggered fermion simulations.  

\vspace*{-0.0cm}

As a conclusion, it would be very useful to 

\vspace*{0.3cm}
\fbox{
\parbox[b]{6.5cm}{
use and test \underline{exact} algorithms\\
and to perform a 
fair cost comparison of exact algorithms including the approach to the
continuum limit. 
}
}
\vspace*{0.3cm}

The actions discussed in section 2 are sometimes very complicated,
including often fattening of links. 
A good news is that methods have been developed that allow for
simulations of such actions either by reject/accept steps
\cite{annalgo} or by computing the force explicitly \cite{kamleh}
using special projection methods. 

As a last topic in this algorithm section I would like to discuss 
a problem of principle. Random matrix theory (RMT) \cite{rmtgeneral}
predicts the  
eigenvalue distribution for the lowest non-vanishing
eigenvalue $\lambda$ in topological charge sector zero to be
$P(\lambda)= \frac{z}{2}e^{-\frac{1}{4}z^2}$ with $z=\lambda\Sigma V$ and
$\Sigma$ the infinite volume scalar condensate
\cite{lowEV}. This prediction  is confirmed
in practical simulations \cite{rmtresults}.  

Such a distribution predicts the occurrence of very small eigenvalues
to be frequent,
and hence close to
the critical point huge fluctuations have to be expected, rendering
the simulations very expensive. 
%Thus, it might be that, although the 
%formula of the algorithmic cost for generating a fixed number 
%of configurations given in 
%eq.~(1) will remain true, many more configurations would have to be 
%generated towards the chiral limit to obtain reliable results.
Of course, a non-vanishing quark
mass will regulate the problem, but it can do so only to a limited
amount when the physical point at realistic values of the quark mass
is reached. The conclusion is therefore {\em that we have to
expect that dynamical fermion simulations are becoming even more  
problematic than the cost formula of eq.~(\ref{algocost}) would imply 
when approaching small quark masses.}
As a result, the need to make contact with chiral perturbation theory
is most urgent. 

\vspace*{-0.1cm}
\section{Scaling in the quenched approximation}

Before turning to any results of unquenched simulations, let me
discuss scaling in the {\em quenched} case first. 
Surprising results were  
presented by Aoki \cite{aokireview} at Lattice2000. 
He used different
actions: staggered fermions with plaquette
action, Wilson and improved Wilson fermions with plaquette 
and renormalization group inspired gauge actions.
His results for $m_N/m_V$ for the different actions as a function
of $m_\mathrm{PS}/m_V$ (i.e., the Edinburgh plot) did not agree 
{\em in the continuum limit} even for {\em for heavy quark masses.}
The staggered results disagreed with the Wilson results,
which agreed among themselves. Surprisingly, there was not
much response to this striking and worrisome observation afterwards. 

\begin{figure}[htb]
%\vspace{9pt}
\includegraphics[width=7.0cm,height=5.8cm]{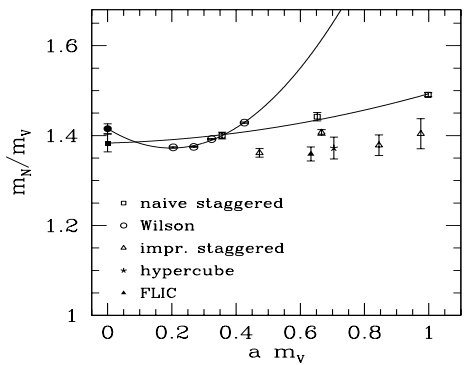}
\vspace*{-0.9cm}
\caption{Continuum extrapolations of unimproved Wilson and 
unimproved staggered fermions, employing a pure $a^2$ dependence  
in the unimproved staggered and an $a+a^2$ dependence in the unimproved 
Wilson case. 
A fixed value of 
$m_{PS}/m_V=0.7$ was chosen. Also included are data for improved staggered, 
hypercube
and FLIC fermions. The data were compiled together with J. Zanotti.}
\label{quenchedscaling}
\end{figure}

The question arises, whether something could be wrong with the
continuum limit of improved staggered fermions. In Aoki's article,
a particular form of the continuum extrapolation was taken, 
linear in the lattice spacing for Wilson type fermions and an
$a^2$ dependence  
for the staggered results. In fig.~\ref{quenchedscaling} I show
an alternative fitting procedure, allowing for an additional 
$a^2$-dependence for the Wilson plaquette data\footnote{I leave out the 
improved Wilson data since they are not very consistent among themselves 
and it would certainly be important to generate a new set of data.}.

\vspace*{-0.2cm}

Although this way of fitting the data might not be the final truth,
it provides a possibility with the conclusion that there is nothing 
wrong with the continuum limit of staggered fermions. The continuum
value for $m_N/m_V$ is also consistent with results for various 
other improved actions as included in the plot.  
It would be very important, to supplement more and better data for such
a plot and to 

\vspace*{0.3cm}
\fbox{
\parbox[b]{6.5cm}{
perform a precise scaling analysis for various fermion actions
in the {\em quenched} approximation. 
}
}
\vspace*{0.3cm}

Such a test would be very re-assuring that all the calculations that 
are done with the many different actions lead to consistent values 
in the continuum limit and can hence lead to a good control over the
systematic errors. For attempts for such a precise scaling tests I
refer to the figure (done together with J. Zanotti) 
in my plenary talk \cite{myplenary} 
and the talks by A. and P. Hasenfratz at the
conference \cite{ahasenfratzscaling,phasenfratzscaling}. 

\vspace*{-0.1cm}
\section{Problems with dynamical fermion simulations}

I now turn to somewhat surprising problems that are already encountered 
in practical simulations 
with dynamical fermions performed as of today. In the light of
these problems, it may not 
be wise to start unthinkingly large scale 
and very expensive simulations with dynamical fermions but rather 
to concentrate on these problems for a while and try to understand
and solve them. 

\vspace*{-0.1cm}
\subsection{Gauge actions}

\begin{figure}[htb]
\vspace{-0.1cm}
\includegraphics[width=7.0cm,height=5.8cm]{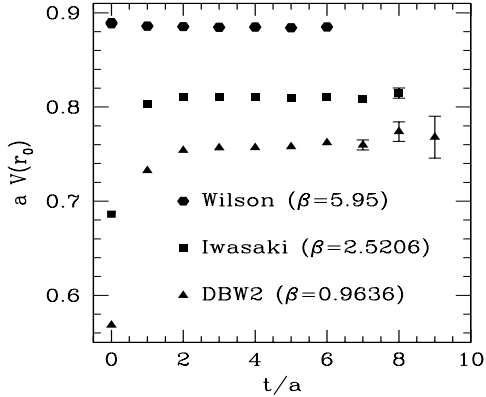}
\vspace*{-0.4cm}
\caption{Static potential for various gauge actions.
Note that, as the effect of complex eigenvalues, 
in the case of improved gauge actions \cite{rgaction} the
potential starts from below.}
\label{potential}
\end{figure}

As mentioned before, improved gauge actions are not reflection
positive. This leads to complex eigenvalues in the transfer matrix
\cite{symanzikgauge}. 
In fig.~\ref{potential} the static potential $V(r_0)$ is shown 
at a distance of
the hadronic scale $r_0$ \cite{rainerscale} as function of the 
euclidian time. The values of $V(r_0)$ for the improved gauge actions approach
their (in time) asymptotic value from below, indicating the effects
of the complex eigenvalues \cite{silvia}. Although for asymptotic
times the value of the potential can be determined, this suggests that 
for these actions larger time extents would have to be used. This 
can have consequences in particular for determinations of glue ball
masses from such actions. 
%The consequences for the determination of the 
%static potential is that as a function of euclidian time,
%$V(r)$ does not approach its value from
%above, but starts from below, showing eventually even an 
%oscillatory behaviour \cite{silvia}.
%\footnote{For an example of the effects of complex eigenvalues on
%correlation functions I refer to ref.~\cite{ghost} where indefinite
%states are introduced. By tuning parameters badly, the correlation
%function starts to oscillate instead of showing a nice exponential
%decay.}.

From a free field analysis \cite{silvia} it also appears that 
different gauge actions might lead to very different lattice artefacts.
In addition, the lattice spacing dependence need not 
to be very smooth approaching the 
continuum limit thus driving a reliable continuum extrapolation difficult.
It might be useful therefore to perform simulations with different
actions to check the continuum limit. 
An example, from my own experience, for such a procedure
can be found in ref.~\cite{universal} where simulations 
using different fermion actions 
were performed to extract 
moments of parton distribution
functions. 
We will come back to this point towards the end of this 
proceedings. 

Another, last difficulty 
of improved gauge actions is that they are very inefficient 
to sample topological charge sectors \cite{topodifficult}.
Although this is not too problematic for quenched simulations since
there many configurations can be generated fast, it may have serious
consequences for dynamical fermions when such actions are employed.

\vspace*{-0.1cm}
\subsection{Wilson fermions}

\begin{figure}[htb]
%\vspace{9pt}
\includegraphics[width=7.0cm,height=5.8cm]{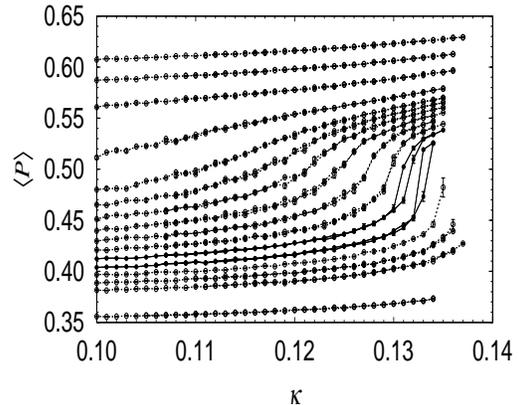}
\vspace*{-0.4cm}
\caption{Thermal cycles in $\kappa$ for the plaquette for 
$N_f=3$ non-perturbatively improved
Wilson fermions at various values of $\beta$ ranging from 
$\beta=4.6$ to $\beta=6$ from bottom to top (from \cite{cppacspd}).
For $\beta\approx 5$ hysteresis effects are observed, indicating 
the existence of a first 
order phase transition.}  
\label{firstorder}
\end{figure}

The CP-PACS collaboration started $N_f=3$ simulations with improved
Wilson fermions and the plaquette gauge action. By performing 
thermal cycles\footnote{In this standard procedure to detect
phase transitions, one starts at some value of $\kappa_\mathrm{start}$,
increases the value of $\kappa$ by some $\Delta\kappa$ until a 
value of $\kappa_\mathrm{final}$ is reached and then runs back to 
$\kappa_\mathrm{start}$.}, clear indications of a first order phase
transitions were detected as can be seen in fig.~\ref{firstorder}
where a pronounced hysteresis effect shows up. 
A similar effect appears also for $N_f=2$ flavours \cite{kjpd}
if again the 
plaquette gauge action and non-perturbatively improved Wilson fermions
\cite{symanzikdynamic} are used.
It might be speculated that these phase transitions are related to
the phase structure of the fundamental-adjoint pure gauge action.
Working in the vicinity of such a phase transition can lead, e.g.,
to unexpectedly
large lattice artefacts that may render the continuum extrapolation
difficult.

Using the Iwasaki action \cite{rgaction} or the Symanzik
gauge action the signs of the first order
phase transitions seem to disappear completely \cite{cppacspd}. 
However, it is not clear,
whether such effects eventually will strike back. Therefore, 
it would be important to 
 
\vspace*{0.2cm}
\fbox{
\parbox[b]{6.5cm}{
study and understand the $T=0$ phase diagram of QCD, to
investigate different gauge actions and to 
understand the nature of the phase transition.
}
}
\vspace*{0.2cm}

\vspace*{-0.2cm}
\begin{figure}[htb]
%\vspace{9pt}
\includegraphics[width=7.0cm,height=5.8cm]{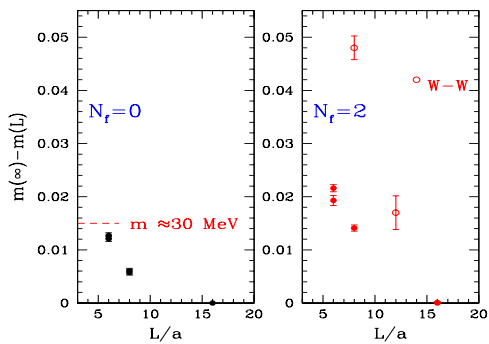}
\vspace*{-0.7cm}
\caption{The quark mass difference of $m(\infty)-m(L)$ in quenched
and in full QCD using non-perturbatively improved Wilson fermions
($L/a=\infty \equiv L/a=16$). Note that for the unquenched case
the lattice artefacts are substantially larger. As an aside, we also show
results from standard Wilson fermions (W-W) where the lattice artefacts are 
even bigger (from \cite{rainerprivate}).} 
\label{quarkmass}
\end{figure}

In fig.~\ref{quarkmass} we show a comparison of the lattice artefacts in
the quark mass for 
quenched and unquenched simulations \cite{rainerprivate} using 
non-perturbatively improved Wilson fermions. 
The figure indicates that for dynamical fermions
the artefacts are substantially larger, although they are still smaller 
than using naive Wilson fermions \cite{sesam}.
Whether this increase of lattice artefacts is due to the possible
phase 
transition discussed above or originates from a completely different source,
is an open question and gives additional 
motivation to understand the phase diagram better.

During the conference it has been pointed out \cite{rainertalk} that
in dynamical simulations using Wilson or staggered fermions
the hadronic scale $r_0$ depends rather strongly on the quark mass
at fixed values of $\beta$.
Clearly, this effect has to be taken into account
when quantities are analyzed as functions of, say, 
$r_0\cdot m_\mathrm{PS}$. This becomes especially important 
when comparisons with chiral perturbation are performed. 
In addition, it was shown that in the 
quantity $r_0(m)/r_0(m_\mathrm{ref})$ with 
$m_\mathrm{ref}$ determined by $m_\mathrm{PS}/m_\mathrm{V}=0.45$, staggered
fermions exhibit surprisingly large lattice artefacts.
Clearly, it would be important to check and clarify this finding 
further with results from improved staggered fermion simulations.

\vspace*{-0.5cm}
\subsection{Staggered fermions}

For simulations with staggered fermions, naive or improved, there 
is a problem with the topological charge identification when 
coarse lattices at large values of the coupling are used. 
This can be seen quite evidently in studies that compare the 
predictions of the eigenvalue distributions predicted by RMT with the results
of numerical staggered fermion simulations. 
Data obtained in topological charge sector
$1$ are fitted with the topology $0$ prediction of RMT, while 
the curves for topology $1$ fail completely
\cite{staggeredtopology}.
These studies were mainly done with unimproved staggered fermions.
For newer results with improved staggered fermions,
see \cite{ahasenfratzscaling}.

In \cite{poulcondpaper} it was argued that this failure of staggered
fermions with respect to questions related to topology 
{\em has to occur} when the gauge fields are not smooth enough and 
one is not working at small enough values of the coupling. 
The reason is that with such rough gauge fields, 
the taste breaking effects are strong and the symmetry breaking
pattern is not the one of the continuum.                           

Indeed, in the first reference of \cite{staggeredtopology} where 
simulations of the Schwinger model have been performed, there are
indications that for larger values of $\beta$ the predictions of RMT
and the numerical data get reconciled. 
Nevertheless, it would be very important

\vspace*{0.3cm}
\fbox{
\parbox[b]{6.5cm}{
to check explicitly
that the predictions 
of RMT and other properties of topology are reproduced by,
preferably, improved staggered fermions. 
}
}
\vspace*{0.3cm}

\vspace*{-0.1cm}
\section{Results for dynamical Wilson fermions}

This section is devoted to some results that have been obtained recently
with dynamical Wilson fermions. 
Results concerning staggered fermions can be
found in \cite{gottliebtalk} and I will only shortly discuss 
staggered fermions
in the context of chiral perturbation theory here.

\vspace*{-0.1cm}
\subsection{Chiral perturbation theory}

The algorithms we are using at the moment will sooner or later
hit a cost wall where the exponential growth of the cost will set in 
with full power 
when the quark mass is lowered below a threshold value.
In order to obtain results for physical values of the pion and $\rho$
masses it becomes therefore mandatory to make contact to chiral
perturbation theory ($\chi$PT) and then use  $\chi$PT to perform the
extrapolations to the ``physical point''. Finding out the window where
on the one hand lattice simulations can be performed and, on the
other hand, $\chi$PT
can be controlled and higher order effects can be safely neglected,
is of central importance in lattice QCD 
and much effort will be invested to resolve this
question. Fig.~1 suggests that improved staggered fermions may reach 
this overlap region sooner than Wilson fermions.

There are two main strategies to confront lattice results with 
predictions of $\chi$PT:

\begin{itemize}
\item the first strategy is to extrapolate 
      lattice data to the continuum limit first 
       and then compare to 
      $\chi$PT;
\item the second strategy is to take the lattice artefacts into account 
      in $\chi$PT itself and compare then to lattice data. 
\end{itemize}

Let me discuss the first strategy first. The big advantage of this procedure
is that in the continuum chiral symmetry is restored and therefore a direct
comparison to $\chi$PT becomes possible\footnote{Of course, using 
chiral invariant lattice fermions would also fulfill this requirement,
but, they are, unfortunately, very expensive to simulate.}.
Clearly, the bad side of the coin is that performing the continuum 
limit is demanding and needs large computer resources. 

This computational demand is at present ``tricked out'' by taking
lattice data at only one value of the lattice spacing. Of course,
the improved lattice actions in both the Wilson and the staggered
cases substantially reduce  
the lattice artefacts. However, only future simulations can 
tell, whether the lattice data were indeed close enough to their continuum
values. 
%In any case, maybe we 
%should confess already now to our friends from the continuum that we
%are doing this ``little trick''. 

In ref.~\cite{bhm} a recent analysis of chiral perturbation theory
applied to lattice data are given. Older discussions are numerous
and I have to refer to \cite{chptdebate}, the plenary discussion
at Lattice 2002 and \cite{chpt} for other examples (taken from 
the analysis of moments of parton distribution functions, my
own hobby-horse).

\begin{figure}[htb]
%\vspace{9pt}
\includegraphics[width=7.0cm,height=5.8cm]{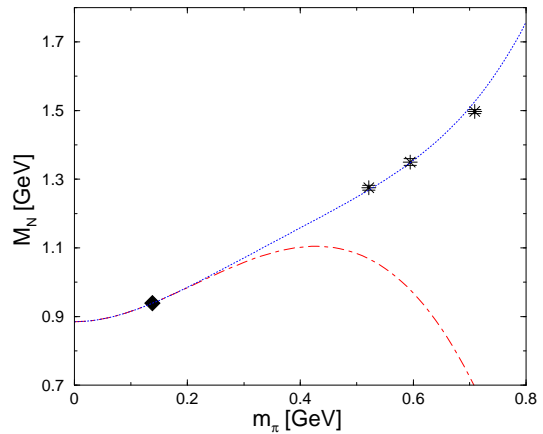}
\vspace*{-0.4cm}
\caption{Comparison of CP-PACS data with different kinds of performing
$\chi$PT (from \cite{bhm}). The dashed dotted curve is a parameter-free third order
result. The dotted line is a fourth order representation using a best fit for the 
additional low energy parameters appearing in the fourth order of $\chi$PT.}
\label{chiralfig1}
\end{figure}

%\begin{figure}[htb]
%\vspace{9pt}
%\includegraphics[width=7.0cm,height=6.0cm]{plots/mass4.ps}
%%\includegraphics[width=16.0cm,height=6.0cm,scale=0.0,angle=0]{plots/pw.ps}
%\caption{Four-loop corrections in $\chi$PT (from \cite{bhm}).}
%\label{chiralfig2}
%\end{figure}
%
In the formulae used to perform the analyzes in 
\cite{bhm}, the parameters of the chiral 
Lagrangian were fixed by taking experimental input values obtained
at the physical 
point. In the calculation chiral symmetry was consistently kept and
an ``improvement term'' to cancel cut-off effects was added. 
The numerical data are taken from 
improved Wilson CP-PACS simulations and are only at one value of the 
lattice spacing. 
Fig.~\ref{chiralfig1} shows a comparison of the predictions from
$\chi$PT in this setup and the data, which seems to be not so bad, 
even for quite large pion masses.
However, in ref.~\cite{bhm} itself it is claimed:
{\em We stress again that applying the expressions to pion masses
above 600 MeV is only done for illustrative purposes, for a realistic
chiral extrapolation smaller pion masses are mandatory.}

The reason for this becomes clear by looking at 
the size of corrections that can appear in a 4-loop
calculation. These corrections can 
be very large, depending on the precise form of performing 
$\chi$PT. I do not have the space
(and it is not the goal of this article)
to discuss in detail which kind of $\chi$PT is correct, what flaws 
there might be and where conceptual difficulties appear. 
However, I believe, that finally it has to be pure $\chi$PT 
that has to be used, without any additional modeling or assumptions.
Only then can we safely employ $\chi$PT to do for us the job of
extrapolating to the physical point.

The second strategy is to take possible lattice artefacts into account
in $\chi$PT itself. This idea was put to the fore by S. Sharpe
\cite{sharpechiral} and was then further developed in 
\cite{chiralwilson} for Wilson and in 
\cite{chiralstaggered} for staggered fermions. 
The advantage of this approach clearly is that chiral symmetry 
violating effects, taste violating effects in the case of staggered
fermions and even different actions can be taken into account 
from the beginning. 
%Another advantage is
%that hybrid calculations become possible where, e.g. standard
%fermion formulations are taken for the sea quark and
%chiral invariant formulations for the valence quark. 
The disadvantage
is that the corresponding chiral Lagrangian has twice as many low energy 
constants, and that the additional parameters
depend on the bare coupling $g_0$ leading to 
a plethora of fit parameters. 

This approach has been used already in practical simulations.
For results with improved staggered fermions I refer to the 
contribution of S. Gottlieb at this conference 
\cite{gottliebtalk}.
The only remark, I want to add here is that $\chi$PT provides 
a justification of taking the square root of the fermion determinant.
This situation corresponds to a partially quenched setup in which two 
quarks are taken to be quenched \cite{bernardgoltermann}. 
The idea can be generalized using the replica trick: take arbitrary
numbers $N_u$, $N_d$, $N_s$ for the $u$, $d$ and $s$  quarks. 
Perform the computation
in this general form and set in the end $N_u=N_d=N_s=1/4$ 
\cite{bernardprivate}. In \cite{ourchiralpt}, 
although in a different context, 
it has been demonstrated
that such a procedure is, at least to 1-loop order,
completely equivalent to the super symmetric
method and, although a proof to all orders is still missing, there
is a good chance that the  
argument might work. 

\begin{figure}[htb]
%\vspace{9pt}
\includegraphics[width=7.0cm,height=5.8cm]{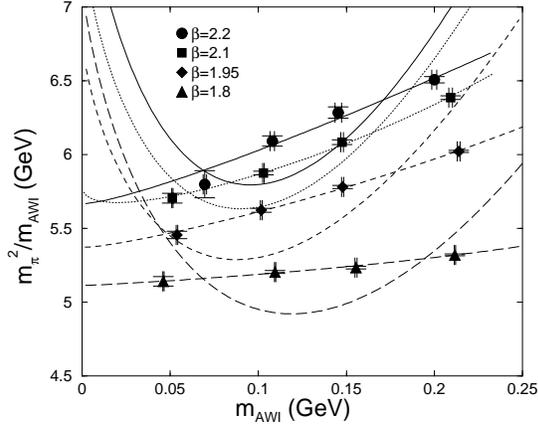}
\vspace*{-0.4cm}
\caption{$m_\mathrm{PS}^2/m_q$ as a function of the quark mass for various
values of $\beta$. 
For each value of $\beta$ two curves are shown. The curves showing
a very strong curvature represent continuum $\chi$PT while the
lines that fit the data represent $\chi$PT with lattice artefacts
taken into account, from \cite{aokichpt}.}
\label{chiralaoki}
\end{figure}

After this excursion to staggered fermions,
let me now turn back to Wilson fermions. 
There are two works at the moment where simulation data are confronted
to the predictions of $\chi$PT when lattice artefacts are included
in $\chi$PT.
The first work is ref.~\cite{aokichpt} and the second is 
ref.~\cite{montvaychpt}.
We show in fig.~\ref{chiralaoki} a plot of $m_\mathrm{PS}^2/m_q$ 
as a function of the quark mass at various values of $\beta$. 
For each value of $\beta$ two curves are shown. The curves showing
a very strong curvature represent continuum $\chi$PT while the
lines that fit the data represent $\chi$PT with lattice artefacts 
taken into account. 
Clearly, continuum $\chi$PT is not able to describe the simulation
data while lattice $\chi$PT fits the data well. 
Another observation is, however, that very large lattice artefacts
in $m_\mathrm{PS}^2/m_q$ seem to exist. Data at various values of
$\beta$ come out to be quite different.

\begin{figure}[htb]
%\vspace{9pt}
\includegraphics[width=7.0cm,height=5.8cm]{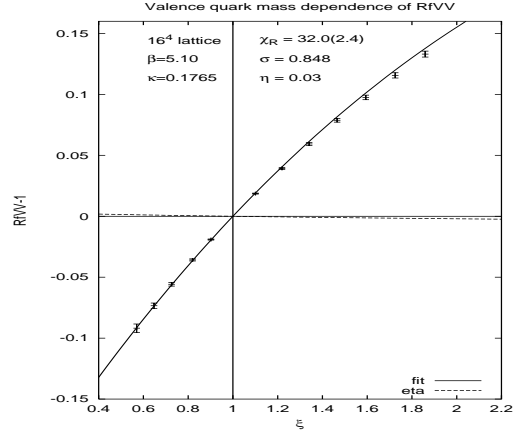}
\vspace*{-0.4cm}
\caption{The double ratio mentioned in the text compared to
$\chi$PT. $\xi$ denotes the ratio of valence to sea quark mass. }
\label{chiralmontvay}
\end{figure}

Another comparison is shown in fig.~\ref{chiralmontvay} (from
\cite{montvaychpt}). Here a ratio of decay constants of the type, 
$f_{VS}^2/(f_{VV}f_{SS})$
is plotted 
with $V$ and $S$ denoting valence and sea quantities.
In this double ratio, which is only one example of the many fits that can be 
found in 
ref.~\cite{montvaychpt}, a good agreement with 
the predictions of $\chi$PT is seen. However, there seem to be
tremendous cancellations of cut-off effects, keeping in mind
that the simulations have been done at a value of the lattice spacing
of $a=0.28$fm.

Another, somewhat worrisome observation is that the universal 
$\Lambda$-parameters come out to be very different, depending 
on the value of $\beta$ that is used (in case of \cite{aokichpt}) 
or how the lattice spacing is defined (in case of \cite{montvaychpt}).
Clearly, we are just in the beginning of using the new approach of
$\chi$PT where lattice spacing effects are taken into account and the 
first results, using this method, are somewhat puzzling. 
In particular, I would like to advocate to de-double the 
double ratios in order to be able to disentangle the cut-off effects
and show their size, furthering thus our understanding of $\chi$PT and the 
connection to lattice QCD. 

\vspace*{-0.1cm}
\section{Numerical results}

In this section, I would like to discuss a few selected results from
recent dynamical Wilson fermion simulations. 

\vspace*{-0.1cm}
\subsection{Finite size effects}

\begin{figure}[htb]
\includegraphics[width=6.5cm,height=5.3cm]{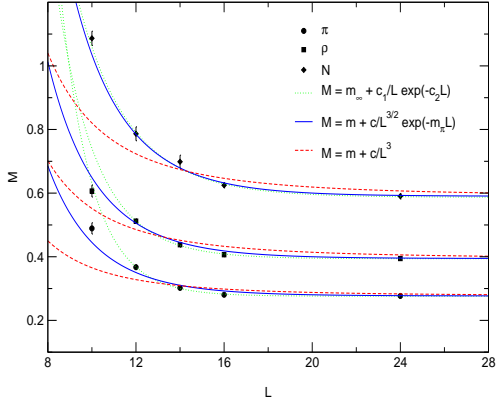}
\vspace*{-0.4cm}
\caption{Finite size effects of the pion, $\rho$ and Nucleon masses
compared to various fits corresponding to different analytical 
predictions (from \cite{sesamfse}). 
The exponential form of finite size effects seem to be clearly
preferred.}
\label{fse}
\end{figure}

It seems that we finally see the 
exponential finite size effects of hadron masses. The general 
form of these finite size effects are
$
M(L)-M = - \frac{3}{16\pi^2 ML}\int_{-\infty}^{\infty} F(iy)
e^{\sqrt{M_\pi^2 + y^2}L}
$
where $F$ is the $\pi-\pi$ forward scattering amplitude in infinite volume
\cite{martinfse}. A calculation of the integral leads to the 
exponential finite size correction $M(L)-M\propto L^{-3/2}e^{-m_\pi L}$.
Corrections to this formula have been computed within 
$\chi$PT \cite{duerrfse}. An alternative calculation was performed
and also tested in ref.~\cite{arifafse}. At this conference there were
two contributions \cite{arifafse,sesamfse}
that showed that indeed the finite size effects
are exponential and not power like of the form
$ M= m_\infty + c/L^3$ as anticipated in ref.~\cite{japanfse}.
In fig.~\ref{fse} a comparison of numerical simulation data to various
fits of the finite size effects are shown. 
It seems that indeed the exponential form of finite size corrections
are preferred by the data confirming thus 
the exponential finite size effects in lattice simulations.
A similar picture and conclusion can be found in \cite{arifafse}.
Clearly, these first results for the finite size effects need to be
confirmed in future simulations.

\vspace{0.3cm}

\vspace*{-0.1cm}
\subsection{Fundamental parameters of QCD}

The running coupling and running quark masses are basic and important
parameters of QCD that can be extracted from lattice simulations. The status
of these calculations were reported at this conference \cite{runningcoupling}.
Although the running itself has been determined already to a quite high
precision, the determination of the physical scale is still missing.
In addition, it would be necessary to complement the present simulations
with even larger lattices to keep the continuum limit under better control.
Both of these goals will certainly be done in the near future.
In the past, 
the Schr\"odinger functional, which was used for the scale dependent
renormalization \cite{schrod}, was the domain of the ALPHA collaboration.
Nowadays, 
there
are new groups who use this renormalization scheme. In particular,
a serious attempt for $N_f=3$ flavours of dynamical fermions  
has been started by the CP-PACS and JLQCD collaborations 
\cite{nf3coupling}. 

The CP-PACS and JLQCD collaborations seem not only to join their forces
to compute the coupling and quark masses but also for other quantities.
This is a convincing indication that dynamical fermions are a real challenge
and a very difficult problem. Present simulations employ the 
Iwasaki gauge action of ref.~\cite{rgaction} combined with the Symanzik improved
Wilson fermion action. The reason for this choice is the occurrence
of a phase transition when the standard gauge action is employed
as mentioned above.
The value of $c_\mathrm{SW}$  
is the non-perturbatively improved one \cite{cppacscsw}.

\begin{figure}[htb]
%\vspace{9pt}
\includegraphics[width=7.0cm,height=3.9cm]{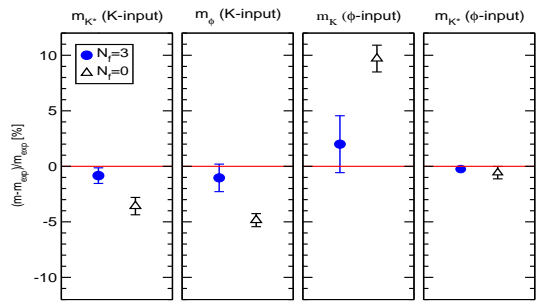}
\vspace*{-0.4cm}
\caption{Relative difference 
of the K and $\Phi$ meson masses as compared to the experimental 
data. Open triangles represent the quenched approximation.}
\label{scaleamb}
\end{figure}

In the $N_f=3$ simulations, the physical extent of the lattice is about
$L\approx 1.6$ fm. With such a lattice, it is not possible to perform
simulations for baryonic quantities reliably since the finite size effects
will be very large \cite{baryonfse}.
Hence, at the moment one concentrates on mesons. 
In fig.~\ref{scaleamb} the relative difference
of the K and $\Phi$ meson masses as compared to the experimental 
data is shown. The open triangles represent the quenched approximation
where for different physical input value a large variation is found.
It is very reassuring that for dynamical fermions this intrinsic 
uncertainty of the quenched approximation seems to be eliminated 
completely. 

Similar to what has been said above for the quenched approximation, it 
would be very interesting to

\vspace*{0.3cm}
\fbox{
\parbox[b]{6.5cm}{
combine Wilson, staggered and other dynamical fermion action
results to check the different approaches in the chiral, continuum and infinite
volume limits.
}
}
\vspace*{0.3cm}

\vspace*{-0.1cm}
\subsection{Other topics}

An article like the present one has to be selective and a number of
topics had to be left out, unfortunately. In the first place, I could 
not discuss progress with dynamical domain wall \cite{dynamicaldwf}
or even overlap \cite{dynamicaloverlap} fermions.
For domain wall fermions it seems that an appropriate gauge action
has been determined where an additional adjoint piece is added 
which seems to solve some problems with unexpected large values
of the residual mass. The simulations for producing physics 
have been started and I am sure that at the next conference we will
see a lot of interesting results.
I would also like to mention some works that discuss the study of
violations of CP \cite{creutz} and even CPT \cite{klinkhamer,susuki} without
being able to go into any detail. 
%how CP can be violated when the quark is chosen to be zero or
%even negative. He finds interesting scenarios and the message
%here in general is that aspects of physics at very small values of the
%quark mass might be rather different from what we know from the
%heavier quark mass values. Of course, the low energy phenomena will
%still be described by chiral perturbation theory at the same time.
%
%In ref.~\cite{klinkhamer} a possible violation of CPT invariance
%for overlap fermions is discussed. It is shown that CPT is violated
%for certain special gauge field configurations. It is not clear, however,
%whether these gauge fields form a set of measure zero and
%are hence not important. Further discussion on this point can be
%found in \cite{susuki}.
%

I could also not go into results for moments of parton distribution
functions from dynamical simulations \cite{dynamicalmoments}, nor 
could I discuss questions concerning topology in depth 
\cite{topology}. Attempts to tackle the notoriously difficult problem
of the $\eta'$ meson by making use of the eigenvalue spectrum
can be found in \cite{etaprime}. My personal opinion here is that
this is an interesting approach, but it still has a number of open
questions that should be addressed in the future. 

\vspace*{-0.1cm}
\section{Conclusion}

%How should I conclude? And what should I recommend?
%Well, one obvious conclusion is to revisit the boxes in the main
%part of this contribution and tackle or solve the problems mentioned there.

Obviously, there are dangerous animals on the lattice 
\cite{tallahasse}. Our present dynamical fermion simulations are still in a
rather early stage but they meet already such dangerous animals:
unexpected phase transitions, questions of field theoretically
non-local actions, actions with unitarity violations, occurrence
of large lattice artefacts, algorithms that become extremely costly
when approaching the physical point, uncertainties in the continuum limit
of different lattice actions, difficulties with the chiral extrapolations.

In the real world, we have learned to tame lions and tigers 
and in lattice field theory this will be done for
sure, too. However, we have to spend work for achieving this: we have to 
explore and understand the phase diagram of (zero temperature) QCD,
we have to study the locality properties 
of taking the square root, we have to explore 
better the phase diagrams of extended gauge actions and understand the
properties of such actions. 
We should perform scaling
tests first in the quenched approximation to demonstrate to ourselves
and to the rest of the world that universality holds (with all the
caveats of the notion of universality in the quenched approximation). 
Participate in algorithm development. It is here, where a breakthrough
can happen which would solve the cost-wall problem we have with existing
algorithms;
it would also be very helpful to join the ILDG \cite{ildg} 
initiative to combine
our efforts and exchange configurations.

One recommendation could be to not put all cards on one particular action.
In fact, it would be wise to first select carefully actions that 
do not show the flaws and problems mentioned in this contribution.
Even after such identifications have been 
made, it may be worthwhile to perform simulations with two (good) 
actions to check for systematic errors. 
%Such 2-actions strategies 
%have been performed in the past in the quenched approximation and resulted
%in a great confidence of the results obtained (see, e.g. \cite{universal}).

Another recommendation could be to use actions tailored for the 
physics problem one is interested in. If problems with rather heavy
quark masses are considered, (theoretically sound) staggered fermions 
or improved Wilson
fermions are a good choice. Going lighter, designer actions such as
FLIC, hypercube, truncated perfect actions or domain wall fermions
with rather small extra dimension might enter the game. 
Finally, for really light quarks, chiral invariant (up to machine 
precision) domain wall and overlap fermions should be used. 

%A last word I would give to the algorithm and machine developers. 
%If you compare the
%algorithm achievements in the, say, last 15 years (which was quite
%remarkable) and the speed up of the machines in the same time, you
%will find that the acceleration from the machine side is orders of
%magnitude better. This emphasis the importance of having a continuous
%progress of computer architectures for lattice gauge theory and we should
%strongly support all the efforts of developing new concepts and ideas
%of putting machines together. Of course, in the end, these machines 
%should be used wisely, which brings us back to the dangerous lattice
%animals that are to be tamed. 

\vspace*{-0.3cm}
\section*{Acknowledgments}
When preparing this talk and delivering it in Tsukuba, I felt like
in front of a minefield \cite{book}. I thank all colleagues 
who helped
me with discussions, suggestions, sending material and pointing out pitfalls 
to cross the minefield without stepping on the mines:
S.~Aoki, C.~Bernard, M.~Creutz, P.~Damgaard, C.~Davies, S.~D\"urr,  
M.~Golterman, S.~Gottlieb, A.~Kronfeld, M.~L\"uscher, I.~Montvay, 
R.~Sommer, T.~Wettig.  

F. Knechtli, S. Gottlieb, M. Hasenbusch and J. Zanotti helped me directly
with the talk by preparing figures or even running some simulations.
I also thank K.I.~Ishikawa for lending me his laptop for the 
time of the conference
and T. Kennedy for giving me his for the presentation. 
Last but not least I want to thank B. Willems for the
minefield picture \cite{myplenary} 
that I would never have been able to draw myself.

\def\NPB #1 #2 #3 {Nucl.~Phys.~{\bf#1} (#2)\ #3}
\def\NPBproc #1 #2 #3 {Nucl.~Phys.~B (Proc. Suppl.) {\bf#1} (#2)\ #3}
\def\PRD #1 #2 #3 {Phys.~Rev.~{\bf#1} (#2)\ #3}
\def\PLB #1 #2 #3 {Phys.~Lett.~{\bf#1} (#2)\ #3}
\def\PRL #1 #2 #3 {Phys.~Rev.~Lett.~{\bf#1} (#2)\ #3}
\def\PR  #1 #2 #3 {Phys.~Rep.~{\bf#1} (#2)\ #3}

\def\etal{{\it et al.}}
\def\ibid{{\it ibid}.}

\end{document}